\documentclass[aps,pra,twocolumn,10pt]{revtex4-1}
\usepackage{amsmath}
\usepackage{latexsym}
\usepackage{graphicx}
\usepackage{amsfonts}
\usepackage{braket}
\usepackage{hyperref}
\usepackage{amssymb}
\usepackage{mathtools} 
\usepackage{nicefrac}
\usepackage[utf8]{inputenc}
\usepackage{dsfont}
\usepackage{colortbl}
\usepackage{xcolor}
\usepackage{pgfmath}
\usepackage{paralist}
\usepackage{nicefrac}
\usepackage{pifont}
\usepackage{ulem}
\hypersetup{
    bookmarks=true,         
    bookmarksopen=true,			
    bookmarksopenlevel=1,		
    unicode=false,          
    pdftoolbar=true,        
    pdfmenubar=true,        
    pdffitwindow=false,     
    pdfstartview={FitH},    
    pdftitle={Gap-independent cooling and hybrid quantum-classical annealing},    
    pdfauthor={Lukas S. Theis},     
    colorlinks=true,       
    linkcolor=blue,          
    citecolor=blue,        
    urlcolor=blue           
}

\pgfmathsetmacro\len{1.0cm/1em}

\newcommand{\comut}[2]{[ #1 , #2 ]} 
\newcommand{\Comut}[2]{\left[ #1 , #2 \right]} 
\newcommand{\op}[1]{#1}
\newcommand{\id}{\op{\mathds{1}}}

\newcommand{\dg}{^\dagger}

\newcommand{\sx}{\op{\sigma}_x}
\newcommand{\sy}{\op{\sigma}_y}
\newcommand{\sz}{\op{\sigma}_z}

\newcommand{\atan}[1]{\mathrm{atan}\left(#1\right)}
\renewcommand{\exp}[1]{\mathrm{exp}\left(#1\right)}

\renewcommand{\coth}[1]{\mathrm{coth}\left(#1\right)}
\renewcommand{\tanh}[1]{\mathrm{tanh}\left(#1\right)}

\newcommand{\del}{\mathrm{\Delta}}

\newcommand{\LR}[1]{\left(#1\right)}
\newcommand{\LRs}[1]{\left[#1\right]}
\newcommand{\abs}[1]{\lvert #1 \rvert}

\newcommand{\PTrace}[2]{\text{Tr}_{#1}\left\{#2\right\}}	

\newcommand{\reffig}[1]{Fig.\,\ref{fig:#1}}
\renewcommand{\refeq}[1]{(\ref{eq:#1})}

\newcommand{\refref}[1]{Ref.\,\cite{#1}}

\definecolor{lukas}{rgb}{0,0,1}
\definecolor{peter}{rgb}{0,1,0}
\definecolor{fwm}{rgb}{0,0.5,0.5}
\definecolor{michael}{rgb}{0.5,0.7,1}

\newcommand{\prlsec}[1]{\textit{#1.---}}

\makeindex

\begin{document}

\title{Gap-independent cooling and hybrid quantum-classical annealing}

\author{L. S. Theis}\email{luk@lusi.uni-sb.de}
\author{Peter K. Schuhmacher}
\author{M. Marthaler}
\author{F. K. Wilhelm}
\affiliation{Theoretical Physics, Saarland University, 66123 Saarbr{\"u}cken, Germany}

\makeatletter
\def\Dated@name{Date: }
\makeatother
\date{\today}

\begin{abstract}
In this letter we present an efficient gap-independent cooling scheme for a quantum annealer that benefits from finite temperatures. We choose a system based on superconducting flux qubits as a prominent example of current quantum annealing platforms. We propose coupling the qubit system transversely to a coplanar waveguide to counter noise and heating that arise from always-present longitudinal thermal noise. We provide a schematic circuit layout for the system and show how, for feasible coupling strengths, we achieve global performance enhancements. Specifically, we achieve cooling improvements of about $50\%$ in the adiabatic and a few hundred percent in the non-adiabatic regime, respectively.
\end{abstract}

\maketitle

\prlsec{Introduction}
Adiabatic Quantum Computation\cite{Albash2018} (AQC) is a promising alternative to the quantum circuit model of computation\cite{Nielsen2009}. The first idea of using adiabatic evolution for solving computational problems appeared in \cite{Apolloni1989} where adiabaticity is used to solve classical combinatorial problems, and was referred to as quantum stochastic optimization. Later on\cite{apolloni1988} the term Quantum Annealing (QA) was introduced. It essentially describes a quantum extension of the classical simulated annealing algorithm\cite{Kirkpatrick1983}, and can natively be implemented in the instruction set of an AQC platform\cite{McGeoch2014}. Similar ideas arose and created terminology such as quantum adiabatic algorithms\cite{Farhi2001} and adiabatic quantum optimization\cite{Smelyanskiy2001}. When the term AQC first appeared\cite{vanDam2001} it was solely focused on optimization but has extended its scope to become an alternate approach to the circuit model over the last years. 

Essentially, in order to solve certain problems using AQC one needs to encode the solution to a given problem in the ground state of a Hamiltonian $\op{H}_1$\cite{Albash2018}. For hard practical problems, this ground state is generally prohibitively slow to reach. Hence, one constructs a Hamiltonian $\op{H}(s)=(1-s)\op{H}_0+s\op{H}_1$ with a fully characterized Hamiltonian $\op{H}_0$ and a parameter $s\in [0,1]$ which represents normalized time. At the beginning of the computation $(s=0)$ the system $\op{H}(0)$ will be prepared in the easily accessible ground state of $\op{H}_0$. Adiabatically changing $s$ from $0$ to $1$ ensures that the Hamiltonian $\op{H}(s)$ will remain in its ground state, and hence, at $s=1$, one can extract the sought ground state of $\op{H}_1$. This approach to quantum computation has been shown\cite{Aharonov2007} to be conceptually as powerful as the quantum circuit model. There are various advantages that make AQC/QA appealing, such as an increased robustness against decoherence\cite{Amin2009} and simpler control. Another downside of the quantum circuit model is the effect of finite temperatures: Generally, one wants to operate at the lowest possible temperature in order to reduce harmful effects originating in non-unitary dynamics\cite{Ashhab2006}. In the context of AQC/QA, however, a thermal environment is expected to be actually helpful\cite{Amin2008a,Arceci2017}.

Nevertheless, there are some downsides that need to be considered when implementing AQC/QA: Perfectly adiabatic sweeps require infinite time. Since, in numerics and experiments, sweep lengths are inevitably finite, there will always be diabatic errors\cite{Chasseur2015}, as can be seen from studies of avoided crossings by means of of Landau-Zener (LZ) physics\cite{Landau1932,Zener1932}. Moreover, although there is evidence for thermally assisted AQC (TA-AQC), it remains a general question how thermal excitations of states close to the ground state can be avoided and/or be reverted efficiently. Since the spectral gap $\del$ between the ground state and the next higher state is generally unknown\cite{Ashhab2006} it remains an important task to find efficient cooling schemes that are independent of $\del$. There exist cooling schemes such as Sisyphos cooling\cite{Grajcar2008} and evaporative cooling\cite{Hauss2008} which require knowledge about the energy gap. 

In this letter we present a cooling scheme that is independent of the energy gap $\del$. Without loss of generality we focus on an annealing platform based on superconducting flux qubits\cite{Orlando1999} and restrict our analysis to the dynamics of dissipative Landau-Zener system. We provide a schematic circuit diagram and a set of quantum master equations that accurately describe the associated spin-boson dynamics of the driven dissipative Landau-Zener system, showing that gap-indendent cooling can be achieved by coupling the qubit transversely to an ohmic environment, in addition to always-present longitudinal thermal noise. Since the effect of the additional transverse coupling can be understood as supplementing the quantum annealing procedure with additional classical annealing, we call the proposed scheme \emph{Hybrid Quantum-Classical Annealing} (HQCA).

\prlsec{Model \& Equations of motion}
\begin{figure}
	\centering
    \includegraphics[width=.95\linewidth]{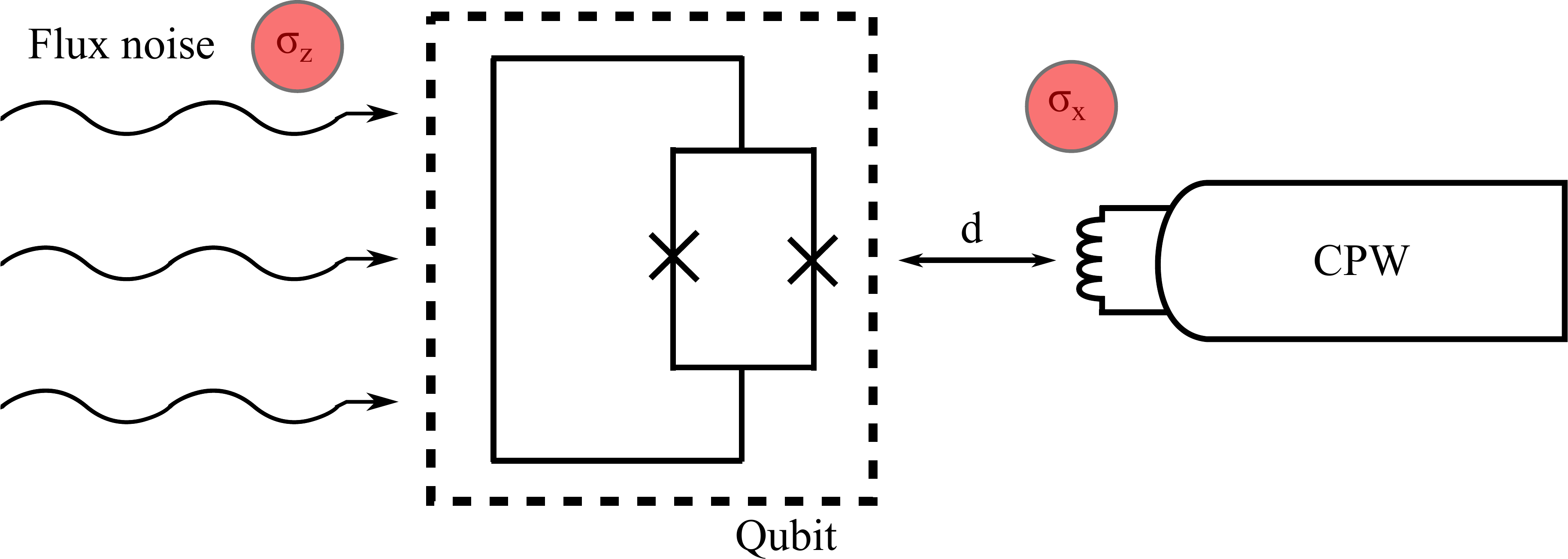}
    \caption{\label{fig:circuit}Schematic circuit diagram to implement both $\sx$ and $\sz$ coupling to a superconducting flux qubit. While the flux noise ($\sz$) is always present, we propose to add an additional $\sx$ coupling in terms of a coplanar waveguide (CPW) at distance $d$ from the qubit. The $\sx$-coupling strength can be controlled directly by altering $d$.}
\end{figure}
As a toy model we restrict ourselves to a dissipative Landau-Zener problem, governed by a spin-boson model\cite{Weiss}. The bare system Hamiltonian $\op{H}_Q(t)$ features a generally time-dependent drive $\epsilon(t)$ and a constant tunneling amplitude $\del$, i.e.
\begin{align}\label{eq:sys_ham}
	\op{H}_Q(t) & = -\frac{\epsilon(t)}{2}\sz-\frac{\del}{2}\sx,
\end{align}
where the $\op{\sigma}_j$ denote the Pauli matrices. In the simplest non-trivial model, $\epsilon(t)$ is a linear in time with sweep velocity $v$ and $y$-intercept $\epsilon_0$, i.e. $\epsilon(t)=vt+\epsilon_0$. Without loss of generality we will assume $\epsilon_0=0$ in the remainder of this letter and let the sweep take place within the time interval $[-t_0,t_0]$ with $t_0$ chosen such that the initial energy splitting is large compared to the gap, i.e. $vt_0=80\del$. This serves as a proper toy model, especially if the two eigenstates can be mapped to well-isolated adiabatic states of a larger system. In fact, a system that features such an isolated small gap has been engineered and analyzed with respect to the influence of (thermal) noise\cite{Dickson2013}. The full Hamiltonian of our system is given by the bare qubit $\op{H}_Q$, the heat bath $\op{H}_B$ and the qubit-environment coupling terms $\op{H}_{QB}$. We model each heat bath as harmonic oscillators and assume that there are both $X$- and $Z$-couplings present, which we will refer to as transverse and longitudinal, respectively. The respective Hamiltonians is then given by
\begin{subequations}\begin{align}
	\op{H}_{QB} & = \sum\limits_{\nu=x,z}\sum\limits_k\op{\sigma}_{\nu}\lambda_{k,\nu}\LR{\op{b}_{k,\nu}+\op{b}_{k,\nu}^{\dagger}},\\
    \op{H}_B & = \sum\limits_{\nu=x,z}\sum\limits_k \omega_{k,\nu}\op{b}_{k,\nu}^{\dagger}\op{b}_{k,\nu}.
\end{align}\end{subequations}
Based on previous ideas and experiments\cite{Novais2005,Kohler2005,Harris2009,Jin2012} we propose a cooling scheme via an additional $\sx$ coupling by using a coplanar waveguide (CPW) as an environment, as shown in \reffig{circuit}. The coupling strength to the qubit can be controlled by modifying the distance $d$ between CPW and qubit. In order to derive an analytic set of equations of motions for the qubit subsystem, we follow the core idea of the standard Bloch-Redfield formalism\cite{Breuer2002}. An adequate model to describe the physics of AQC/QA is the spin-boson model\cite{Weiss}, which properly characterizes the coupling of some quantum system with an external environment. In order to obtain analytic expressions for the equations of motion in case of generic time-dependent Hamiltonians we apply an appropriate formulation\cite{Nalbach2014,Yamaguchi2017} of the Bloch-Redfield theory. Following Refs.\cite{Nalbach2014,Arceci2017} we transform to a frame defined by the time-dependent rotation $\op{R}(t)=\exp{i\phi(t)\sy/2}$ and denote operators in that frame with a tilde, i.e. $\op{\tilde{O}}(t)=\op{R}(t)\op{O}(t)\op{R}^{\dagger}(t)$. Since the transformation is time-dependent the qubit Hamiltonian acquires an additional inertial term, which can be related to non-stoquastic interactions in a multi-qubit scenario \cite{Vinci2017}, so that the Landau-Zener Hamiltonian in the rotating frame reads
\begin{align}\begin{split}\label{eq:qubit_sys}
	\op{\tilde{H}}_Q(t) & = -\frac{E(t)}{2}\sx + \frac{\dot{\phi}(t)}{2}\sy
\end{split}\end{align}
where we use the mixing angle $\phi(t)=\mathrm{atan}(\epsilon(t)/\del)$ and the instantaneous energy splitting $E(t)=\sqrt{\del^2+\epsilon^2(t)}$. For later use we define $\op{\tilde{H}}_0(t) \equiv -E(t)\sx/2$. Analogously, the qubit-environment coupling becomes
\begin{align}
	\op{\tilde{H}}_{QB}(t) & = \sum\limits_{\nu=x,z}\sum\limits_k\op{\tilde{\sigma}}_{\nu}(t)\lambda_{k,\nu}\LR{\op{b}_{k,\nu}+\op{b}_{k,\nu}^{\dagger}}
\end{align}
with $\op{\tilde{\sigma}}_{\nu}(t)$ being the Pauli matrices in the rotating frame. By introducing the weights $f_1(t)=\mathrm{sin}(\phi(t))$ and $f_2(t)=\mathrm{cos}(\phi(t))$ we can express the rotating-frame-matrices as $\op{\tilde{\sigma}}_x(t)=-f_1(t)\sz+f_2(t)\sx$ and $\op{\tilde{\sigma}}_z(t)=f_2(t)\sz+f_1(t)\sx$, respectively.  In order to provide closed analytical expressions for the equations of motion, one employs standard Markovian approximations and an additional adiabatic-Markovian approximation\cite{Nalbach2014} (AMA). The latter is inevitable to deal with the interaction picture transformation needed to carry out the time-dependent Bloch-Redfield formalism. For a detailed derivation, please see Appendix \ref{sec:app_A}. The AMA features two important parts: (i) the memory time of the bath $\tau_{\rm mem}$ is assumed to be much smaller than any system time scale and (ii) the drive $\epsilon(t)$ approximately acts on time scales much larger than $\tau_{\rm mem}$ so that it has no significant contribution to the rates. This, in turn, allows to derive the Bloch equations for the density matrix $\op{\tilde{\rho}}_Q(t)=(\id+\sum_nr_n(t)\op{\sigma}_n)/2$ associated to the qubit subsystem \refeq{qubit_sys}. The Bloch vector $(r_x,r_y,r_z)$ is determined by the set of quantum master equations (QME)
\begin{subequations}\label{eq:bloch_eqs}\begin{align}
	\dot{r}_x & = \LR{\dot{\phi}-\gamma_{xz}}r_z - \gamma_r\LR{r_x-\bar{r}_x},\\
    \dot{r}_y & = E_tr_z - \LR{\gamma_d+\gamma_r}r_y,\\
    \dot{r}_z & = -\dot{\phi}r_x-E_tr_y-\gamma_dr_z-\gamma_{zx}\LR{r_x-\bar{r}_x}.
\end{align}\end{subequations}
Here, we use the shorthand notation $E_t\equiv E(t)$, $\bar{r}_x\equiv\tanh{\beta E_t/2}$ and defined the set of rates
\begin{subequations}\label{eq:rates}\begin{align}
	\gamma_r & = 2\pi\coth{\frac{\beta E_t}{2}}\LR{f_1^2J_x(E_t)+f_2^2J_z(E_t)},\label{eq:rate_relax}\\
    \gamma_d & = 4\pi\lim\limits_{\omega\to 0}\bar{n}(\omega)\LR{J_z(\omega)+J_x(\omega)},\label{eq:limit_1}\\
    \gamma_{xz} & = 4\pi f_1f_2\lim\limits_{\omega\to 0}\bar{n}(\omega)\LR{J_x(\omega)-J_z(\omega)},\label{eq:limit_2}\\
    \gamma_{zx} & = 2\pi f_1f_2\coth{\frac{\beta E_t}{2}}\LR{J_x(E_t)-J_z(E_t)},
\end{align}\end{subequations}
that depend on the spectral densities $J_{\nu}(\omega)$ of the respective environments. Relaxation is encoded in $\gamma_r$, while $\gamma_d$ and $\gamma_{zx,xz}$ describe pure dephasing and cross-dephasing, respectively. We stress that the Bloch-type equations \refeq{bloch_eqs} are based on a proper treatment of external drives. The performed AMA might suggest that the QME are only valid inside the adiabatic regime, i.e. when $v\ll\del^2$. However, even for non-adiabatic drives they are still a good approximation. This has been verified numerically for a similar Hamiltonian in \refref{Arceci2017} by comparing the numerical solutions of their equivalent of Eqs.\refeq{bloch_eqs} to a numerically exact solution obtained via the path integral based method QUAPI\cite{Makri1995}. Furthermore, a detailed analysis of the assumptions that lead to the QME in terms of different time scales has been carried out in \refref{Yamaguchi2017}.

\begin{figure}
	\centering
    \includegraphics[width=.96\linewidth]{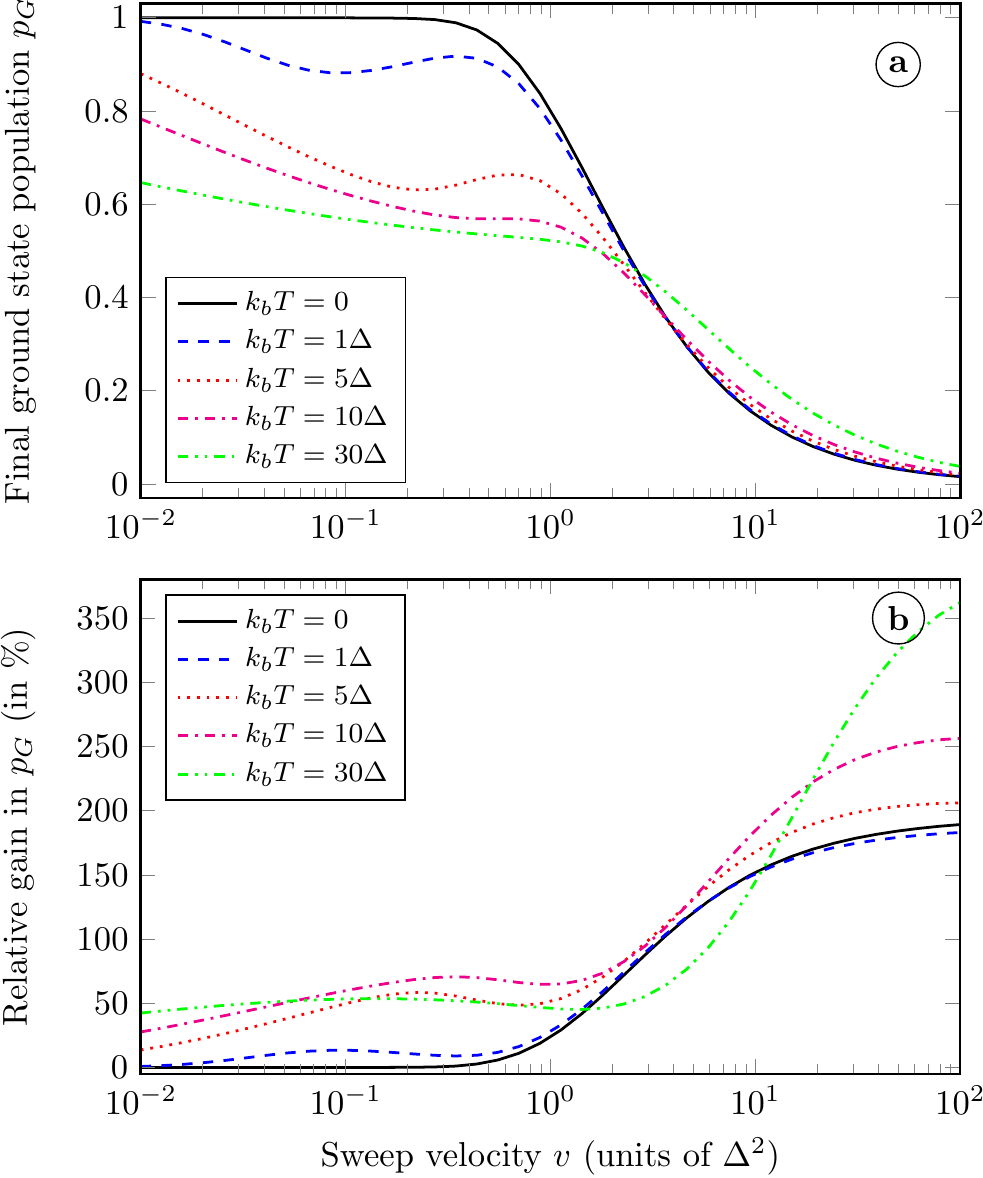}
	\caption{\label{fig:fig_pg}(a) Final ground state population $p_G$ as a function of the sweep velocity $v$ for a $\sz$-only coupling with coupling strength $\alpha_z=5\cdot 10^{-3}$ at different temperatures. Clearly, even for small velocities and small temperatures, a significant amount of population is lost into the excited state owing to heating. (b) Relative improvement of $p_G$ compared to the data in (a) if an additional CPW is used as an additional $\sx$ heat bath with coupling strength $\alpha_x=\alpha_z$, i.e. we plot $(p_G^{(x,z)}-p_G^{(z)})/p_G^{(z)}$ where the superscript indicates the type of couplings in the system. In the adiabatic regime we find improvements of about $50\%$ while the cooling effect in the non-adiabatic regime is even more pronounced with gains of a few hundred percent. Generally, the gain increases with temperature -- indicating proper TA-AQC.}
\end{figure}

\prlsec{Environmental engineering}
In our analysis we restrict ourselves to the case of ohmic heat baths\cite{Shnirman2002,Leggett1987}. That is, the spectral densities $J_{\nu}(\omega)$ depend linearly on $\omega$. However, this model is only valid up to some high-frequency cutoff $\omega_{c,\nu}$. For our purpose, we choose to work with an exponential cutoff at frequencies $\omega_{c,\nu}=10\del$ whereby the exact numerical value has an irrelevant impact on the quality of our results. Different coupling strengths are modeled by the parameter $\alpha_{\nu}$, so that the spectral density is eventually given by $J_{\nu}(\omega)=\alpha_{\nu}\omega e^{-\omega/\omega_{c,\nu}}$. With this explicit form of $J(\omega)$ we compute the limit $\lim_{\omega\to 0}\bar{n}(\omega)J_{\nu}(\omega)$ needed in Eqs.\refeq{rates} to be equal to $\alpha_{\nu}/\beta$. We simulate the QME \refeq{bloch_eqs} with initial conditions set up such that the system will always start in the exact ground state of Hamiltonian \refeq{qubit_sys}. We use the final ground state population $p_G$ after a full Landau-Zener sweep as our figure of merit to evaluate cooling effects. 

In \reffig{fig_pg}(a) we depict the dependence of $p_G$ on the sweep velocity $v$, temperature $T$ and for a pure $\sz$ coupling with $\alpha_z=5\cdot 10^{-3}$. As one expects, thermal excitations heat the system significantly, leading to significant population loss compared to coherent dynamics. If temperatures are not too high, i.e. $k_BT\lesssim 5\del$, there is a locally optimal velocity $v_0$ at which the sum of diabatic errors due to finite sweep length and thermal excitations are minimized\cite{Keck2017}. However, since both, $v_0$ and $p_G(v_0)$, strongly depend on $\alpha_z$ and temperature, sweeping with velocity $v_0$ would be a tradeoff which still features poor performance. Instead, we deduce from \reffig{fig_pg}(b) that an additional CPW coupled via $\sx$ with $\alpha_x=\alpha_z$ generally performs significantly better compared to the situation with only longitudinal thermal noise. The relative gain is defined as $(p_G^{(x,z)}-p_G^{(z)})/p_G^{(z)}$ where the superscript indicates the type of couplings in the system. Moreover, we find that -- except for a small subset of velocities -- higher temperatures lead to better results than low-temperature simulations. We therefore argue that an additional transversely coupled heat bath not only reduces heating -- it also properly demonstrates TA-AQC\cite{Amin2008a}: the benefit of a thermal environment during open system dynamics. We observe this effect even for $\alpha_z > \alpha_x$, remarking that it is slightly attenuated compared to the situation $\alpha_z \leq \alpha_x$. Aside, we remark that the results for higher temperatures serve as a mock-up for small energy gaps.

\begin{figure}
	\centering
    \includegraphics[width=.99\linewidth]{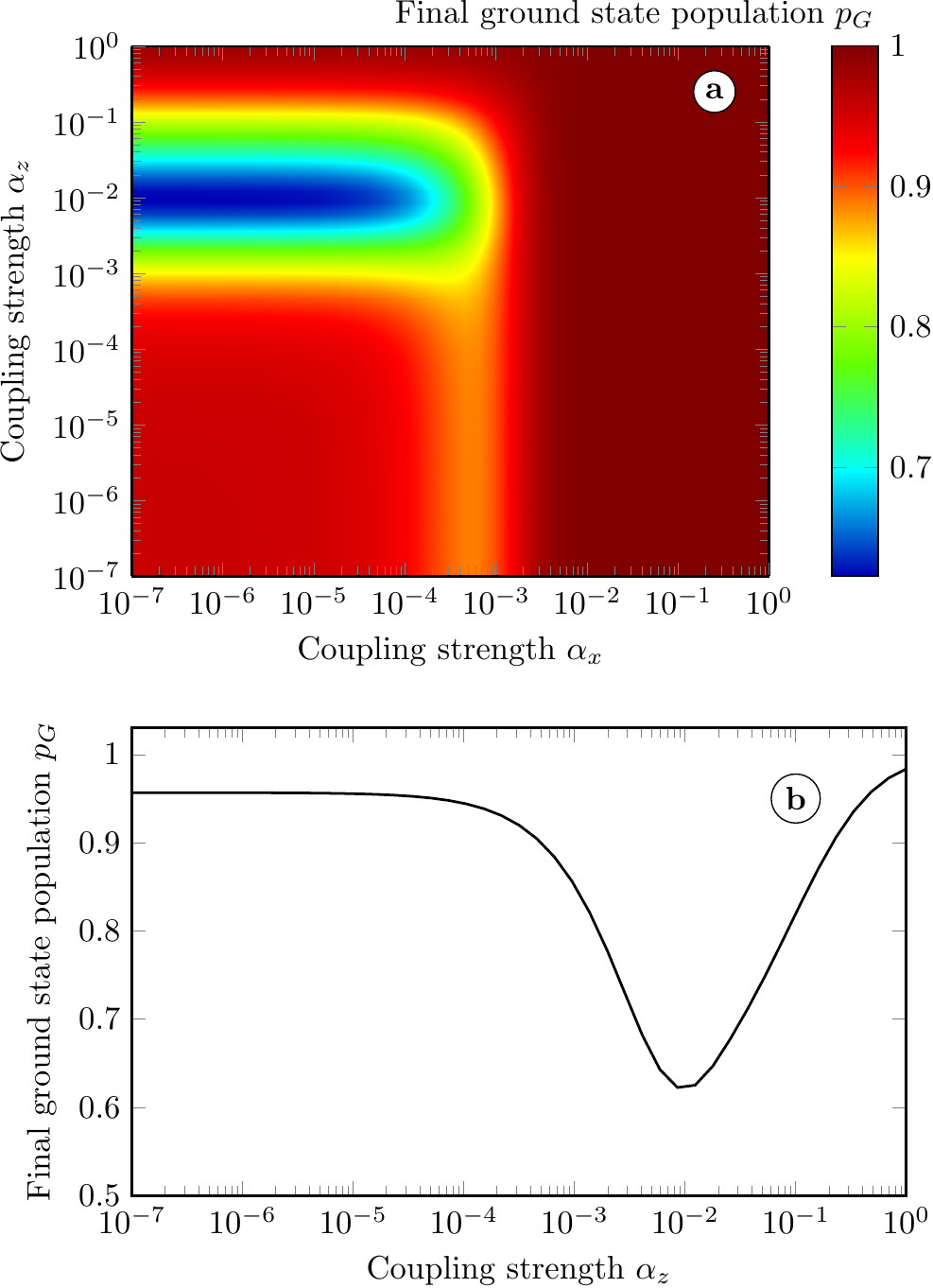}
	\caption{\label{fig:fig_couplings}(a) Dependence of the final ground state population $p_G$ on the coupling strengths $\alpha_x$ and $\alpha_z$ for a temperature of $k_BT=5\del$ at a sweep velocity $v=0.5\del^2$. The velocity is chosen such that it corresponds to a local optimum of $p_G(v)$ as extracted from \reffig{fig_pg}(a). (b) Dependence of $p_G$ on $\alpha_z$ without the existence of an additional CPW, i.e. for $\alpha_x=0$, with identical parameters as in (a). The minimum is reached at $\alpha_{z,0}\approx 0.01$.}
\end{figure}

In case of pure thermal noise ($\sz$), we only observe negligible TA-AQC for reasonable values of $\alpha_z$ in the non-adiabatic regime. Nevertheless, for $\alpha_z \gtrsim \mathcal{O}(0.01)$, we find appreciable indications for TA-AQC even without an additional CPW. A detailed numerical study of how the final ground state population depends on $\alpha_x$ and $\alpha_z$ for fixed temperature $k_BT=5\del$ and fixed velocity $v=0.5\del^2$ is depicted in \reffig{fig_couplings}(a). Comparing to the behavior of $p_G(\alpha_z)$ as shown in \reffig{fig_couplings}(b) the advantage of an additional $\sx$ heat bath becomes apparent: as soon as even a small coupling $\alpha_x$ is present, pronounced relaxation after sweeping through the avoided crossing leads to significant cooling of the system. This is apparent from Eq.\refeq{rate_relax}: Contributions to the relaxation rate $\gamma_r$ are non-negative so that additional transverse coupling amplifies relaxation processes. 

Based on the concept of frustrated decoherence\cite{Kohler2005,Novais2005} one might suspect that excitations into the excited state are effectively blocked due to the non-commutativity of $\sx$ and $\sz$. However, we do not observe such quantum effects (which are similar to the Zeno blockade\cite{Kraus1981}) and attribute the efficiency of the cooling scheme solely to enhanced relaxation effects, as illustrated in Appendix \ref{sec:app_B}. Hence, the general quantum annealing process is supported by relaxation processes at finite temperatures that must be smaller than $E(t)$ well outside the avoided crossing regime; which is similar to the classical simulated annealing\cite{Kirkpatrick1983} algorithm. We therefore refer to our method as \emph{Hybrid Quantum-Classical Annealing} (HQCA). 

If the transverse coupling exceeds $\alpha_x \gtrsim 5\cdot 10^{-3}$, roughly all population has relaxed back to the ground state by the end of the sweep -- irrespective of $\alpha_z$. The value $\alpha_{z,0}$ where the curve $p_G(\alpha_z)$ reaches its minimum decreases with increasing temperature. Note that the non-monotonic behavior of $p_G(\alpha_z)$ that is shown in \reffig{fig_couplings}(b) can be explained using a key result of \refref{Amin2008b}, where the authors show how dissipative dynamics merge into semiclassical dynamics if the associated rates exceed a certain temperature-dependent value. In that case, the final ground state population will be approximately given by the result of coherent dynamics -- which can be estimated via the Landau-Zener formula\cite{Landau1932,Zener1932} $p_G^{\rm LZ}=1-e^{-\pi\del^2/(2v)}$. For the parameters in \reffig{fig_couplings} this corresponds to a semiclassical limit of about $0.95$, which is in good agreement to the curve in \reffig{fig_couplings}(b) for $\alpha_z\sim 1$. 

\prlsec{Conclusion}
In conclusion, we presented a gap-independent cooling scheme for a quantum system affected by $\sz$ noise. Our method generally increases the ground state population after sweeping through an avoided crossing at finite temperatures, owing to enhanced relaxation processes induced by an additional transversely coupled heat bath in form of a coplanar waveguide. We find numerical evidence for significant effects of thermally assisted quantum annealing, and numerically demonstrated that the proposed cooling scheme improves ground state populations by up to a few hundred percent. Thereby we developed a method that has potential to improve the quality of current quantum annealing devices. Recall that parameters are independent of the energy gap, so that the cooling scheme is intrinsically robust against fluctuations of the energy gap.

For further details on the derivation of the QME \refeq{bloch_eqs} and numerical details that illustrate enhanced relaxation we refer the reader to appendices \ref{sec:app_A} and \ref{sec:app_B}, respectively.

\prlsec{Acknowledgements}
The research is based upon work (partially) supported by the Office of the Director of National Intelligence (ODNI), Intelligence Advanced Research Projects Activity (IARPA), via the U.S. Army Research Office contract W911NF-17-C-0050. The views and conclusions contained herein are those of the authors and should not be interpreted as necessarily representing the official policies or endorsements, either expressed or implied, of the ODNI, IARPA, or the U.S. Government. The U.S. Government is authorized to reproduce and distribute reprints for Governmental purposes notwithstanding any copyright annotation thereon.

\bibliography{Bibliography.bib}
\bibliographystyle{apsrev4-1}

\newpage
\onecolumngrid
\appendix
\section{Derivation of the quantum master equations\label{sec:app_A}}
We provide details on the derivation of the quantum master equation. The total Hamiltonian is decomposed as
\begin{align}
	\op{H}(t) & = \op{H}_Q(t) + \op{H}_{QB} + \op{H}_B, \text{ with}\\
    \op{H}_Q(t)  & = -\frac{\epsilon(t)}{2}\sz-\frac{\del}{2}\sx\\
    \op{H}_{QB} & = \sum\limits_{\nu=x,z}\sum\limits_k\op{\sigma}_{\nu}\lambda_{k,\nu}\LR{\op{b}_{k,\nu}+\op{b}_{k,\nu}^{\dagger}},\\
    \op{H}_B  & = \sum\limits_{\nu=x,z}\sum\limits_k \omega_{k,\nu}\op{b}_{k,\nu}^{\dagger}\op{b}_{k,\nu}.
\end{align}
Following \refref{Nalbach2014}, we move to the rotating frame defined by the transformation $\op{R}(t)=\exp{\frac{i}{2}\phi(t)\sy}$ with $\phi(t)=\atan{\epsilon(t)/\del}$. With the instantaneous energy splitting $E(t)=\sqrt{\epsilon^2(t)+\del^2}$ the bare system Hamiltonian and the coupling term become
\begin{align}
	\op{\tilde{H}}_Q(t) & = -\frac{E(t)}{2}\sx + \frac{\dot{\phi}(t)}{2}\sy \equiv \op{\tilde{H}}_0 + \frac{\dot{\phi}(t)}{2}\sy, \\
    \op{\tilde{H}}_{QB}(t) & = \sum\limits_{\nu=x,z}\sum\limits_k\op{\tilde{\sigma}}_{\nu}(t)\lambda_{k,\nu}\LR{\op{b}_{k,\nu}+\op{b}_{k,\nu}^{\dagger}}.
\end{align}
By introducing the weights $f_1(t)=\mathrm{sin}(\phi(t))$ and $f_2(t)=\mathrm{cos}(\phi(t))$ we express the rotating-frame Pauli matrices as 
\begin{align}
	\op{\tilde{\sigma}}_x(t)  =-f_1(t)\sz+f_2(t)\sx, \quad \op{\tilde{\sigma}}_z(t)  =f_2(t)\sz+f_1(t)\sx.
\end{align}
Following standard Bloch-Redfield theory (cf. Sec 3.3 in \cite{Breuer2002}) we start in the interaction frame with respect to $\op{\tilde{H}}_Q$ and $\op{\tilde{H}}_B$. Hence, the coupling Hamiltonian in the interaction picture is given by
\begin{align}
	\op{\tilde{H}}_{QB,I}(t)  = \sum\limits_{\nu=x,z}\op{\tilde{U}}_Q(t)\op{\tilde{\sigma}}_{\nu}(t)\op{\tilde{U}}_Q\dg(t)\otimes\op{B}_{\nu}(t), \quad \op{B}_{\nu}(t) = \sum\limits_k \lambda_{k,\nu}\LR{e^{i\omega_{k,\nu}t}\op{b}_{k,\nu}\dg + e^{-i\omega_{k,\nu}t}\op{b}_{k,\nu}}
\end{align}
with some bath operator $\op{B}_{\nu}$ and the free propagator of the bare qubit $\op{\tilde{U}}_Q(t) = \op{\mathcal{T}}\exp{-i\int_0^t\op{\tilde{H}}_Q(t')\mathrm{d}t'}$. The equation of motion for the density matrix of the reduced qubit subsystem is hence given by
\begin{align}
	\dot{\op{\tilde{\rho}}}_{Q,I}(t) & = -\int\limits_0^{\infty}\mathrm{d}s\,\PTrace{B}{\Comut{\op{\tilde{H}}_{QB,I}(t)}{\Comut{\op{\tilde{H}}_{QB,I}(t-s)}{\op{\tilde{\rho}}_{Q,I}(t)\otimes\op{\rho}_B}}}\\
    & = -\int\limits_0^{\infty}\mathrm{d}s\,\sum\limits_{\nu,\nu'}\left\{\op{\tilde{\sigma}}_{\nu,I}(t)\op{\tilde{\sigma}}_{\nu',I}(t-s)\op{\tilde{\rho}}_{Q,I}(t)\braket{\op{B}_{\nu}(t)\op{B}_{\nu'}(t-s)}\right.\notag\\
    & \hphantom{ = -\int\limits_0^t\mathrm{d}s\,\sum\limits_{\nu,\nu'}(}\left.-\op{\tilde{\sigma}}_{\nu,I}(t)\op{\tilde{\rho}}_{Q,I}(t)\op{\tilde{\sigma}}_{\nu',I}(t-s)\braket{\op{B}_{\nu'}(t-s)\op{B}_{\nu}(t)}+\mathrm{h.c.}\right\}\label{eq:integro-dgl-1}
\end{align}
In the above equation we have already included (i) a weak-coupling approximation (Born approximation), which states that the reservoir is negligibly affected by the system so that we may write the full density matrix as a tensor product $\op{\tilde{\rho}}(t) = \op{\tilde{\rho}}_Q(t)\otimes\op{\rho}_B$ and (ii) a Markovian approximation. The latter states that there is no memory, i.e. time evolution of the state depends only on its present value, and is based on the assumption that the correlation functions decay sufficiently fast compared to the time scale over which the system changes notably. If we choose $\op{\rho}_B$ to be a stationary state of the reservoir, the correlation functions are homogeneous in time, hence $\braket{\op{B}_{\alpha}(t)\op{B}_{\beta}(t-s)}=\braket{\op{B}_{\alpha}(s)\op{B}_{\beta}(0)}$. Furthermore we assume that there is no correlation between different baths, i.e. $\braket{\op{B}_{\alpha}(s)\op{B}_{\beta}(0)} \propto \delta_{\alpha\beta}$. We can then write Eq.\eqref{eq:integro-dgl-1}  in the form
\begin{align}
	\dot{\op{\tilde{\rho}}}_{Q,I}(t) & = -\int\limits_0^{\infty}\mathrm{d}s\,\sum\limits_{\nu=x,z}\left\{\Comut{\op{\tilde{\sigma}}_{\nu,I}(t)}{\op{\tilde{\sigma}}_{\nu,I}(t-s)\op{\tilde{\rho}}_{Q,I}(t)}\braket{\op{B}_{\nu}(s)\op{B}_{\nu}(0)} + \mathrm{h.c.} \right\}.
\end{align}
We are looking for the equation of motion in the Schr\"odinger picture, that is the evolution of $\op{\tilde{\rho}}_Q(t)$, which we obtain by computing $\dot{\op{\tilde{\rho}}}_Q(t) = \op{\tilde{U}}_Q(t)\dot{\op{\tilde{\rho}}}_{Q,I}(t)\op{\tilde{U}}_Q\dg(t)-i\comut{\op{\tilde{H}}_Q(t)}{\op{\tilde{\rho}}_Q(t)}$. A straightforward calculation reveals the sought equation of motion in the Schr\"odinger picture to be 
\begin{align}\label{eq:dgl_1}
	\dot{\op{\tilde{\rho}}}_Q(t) & = -i\Comut{\op{\tilde{H}}_Q(t)}{\op{\tilde{\rho}}_Q(t)} - \sum\limits_{\nu=x,z}\left\{\Comut{\op{\tilde{\sigma}}_{\nu}(t)}{\op{\tilde{S}}_{\nu}(t)\op{\tilde{\rho}}_Q(t)} + \mathrm{h.c.}\right\}
\end{align}
where we introduced the operator 
\begin{align}\label{eq:Sop_1}
	\op{\tilde{S}}_{\nu}(t) & = \int\limits_0^{\infty}\mathrm{d}s\,\op{\tilde{U}}_Q(t,t-s)\op{\tilde{\sigma}}_{\nu}(t-s)\op{\tilde{U}}_Q\dg(t,t-s)\braket{\op{B}_{\nu}(s)\op{B}_{\nu}(0)}.
\end{align}
In order to derive an analytic form for the equation of motion we further need to apply an adiabatic Markovian approximation\cite{Nalbach2014} which amounts to expressing the propagator as 
\begin{align}
	\op{\tilde{U}}_Q(t,t-s) & \approx \exp{-i\op{\tilde{H}}_Q(t)s}.
\end{align}
This is sufficiently accurate provided the memory time $\tau_{\rm mem}$ of the bath is much smaller than any system time scale, $\tau_{\rm mem} \ll (t-s)$, and if the drive $\epsilon(t)$ acts on time scales $\tau_{\epsilon} \gg \tau_{\rm mem}$ so that it has no significant effect on the rates. The correlation function can be expressed in terms of the spectral density $J_{\nu}(\omega)$ of the bath (cf Sec. 3.1.4 in \cite{Weiss}):
\begin{align}\label{eq:corr_1}
		\braket{\op{B}_{\nu}(s)\op{B}_{\nu}(0)} & = \int\limits_0^{\infty}\mathrm{d}\omega\,J_{\nu}(\omega)\LRs{e^{-i\omega t}\LR{\bar{n}_{\nu}(\omega)+1} + e^{i\omega t}\bar{n}_{\nu}(\omega)}
\end{align}
with the single-particle Bose distribution $\bar{n}_{\nu}(\omega)=1/(e^{\beta_{\nu}\omega}-1)$. Using the identity $\bar{n}_{\nu}(-\omega)=-(\bar{n}_{\nu}(\omega)+1)$ we can rewrite Eq.\eqref{eq:corr_1} as an integral over positive and negative $\omega$, i.e.
\begin{align}\label{eq:corr_2}
	\braket{\op{B}_{\nu}(s)\op{B}_{\nu}(0)} & = \int\limits_{-\infty}^{\infty}\mathrm{d}\omega\,\mathrm{sgn}(\omega)J_{\nu}(\abs{\omega})\bar{n}_{\nu}(\omega)e^{i\omega t}.
\end{align}
Inserting Eq.\eqref{eq:corr_2} into the definition \eqref{eq:Sop_1} allows us to carry out the integration over $s$ first, which yields terms $\int_0^{\infty}\mathrm{d}se^{i\omega s}\approx\pi\delta(\omega)$. Note that we here neglect imaginary parts resulting from principal value integrals since they simply manifest themselves as Lamb shifts. Calculating the right hand side of Eq.\eqref{eq:dgl_1} while using the Bloch representation $\op{\tilde{\rho}}_Q(t)=(\id+\sum_nr_n(t)\op{\sigma}_n)/2$ we eventually find the quantum master equations presented in the main article,
\begin{subequations}\label{eq:bloch_eqs}\begin{align}
	\dot{r}_x & = \LR{\dot{\phi}-\gamma_{xz}}r_z - \gamma_r\LR{r_x-\bar{r}_x},\\
    \dot{r}_y & = E_tr_z - \LR{\gamma_d+\gamma_r}r_y,\\
    \dot{r}_z & = -\dot{\phi}r_x-E_tr_y-\gamma_dr_z-\gamma_{zx}\LR{r_x-\bar{r}_x}.
\end{align}\end{subequations}
Here, we use the shorthand notation $E_t\equiv E(t)$, $\bar{r}_x\equiv\tanh{\beta E_t/2}$. Note that we assume same temperatures for both baths since, in experiments, they will both be located in the same cyrostat. The rates are then given by
\begin{alignat*}{2}
    \gamma_r & = 2\pi\coth{\frac{\beta E_t}{2}}\LR{\mathrm{sin}^2(\phi)J_x(E_t)+\mathrm{cos}^2(\phi)J_z(E_t)}, \quad & \gamma_d &= 4\pi\lim\limits_{\omega\to 0}\bar{n}(\omega)\LR{J_z(\omega)+J_x(\omega)},\\ 
    \gamma_{xz} & = 4\pi \mathrm{sin}(\phi)\mathrm{cos}(\phi)\lim\limits_{\omega\to 0}\bar{n}(\omega)\LR{J_x(\omega)-J_z(\omega)}, \quad & \gamma_{zx} &= 2\pi \mathrm{sin}(\phi)\mathrm{cos}(\phi)\coth{\frac{\beta E_t}{2}}\LR{J_x(E_t)-J_z(E_t)}. 
\end{alignat*}

\section{Numerical verification of relaxation and cooling\label{sec:app_B}}
\begin{figure}
	\centering
    \includegraphics[width=.8\textwidth]{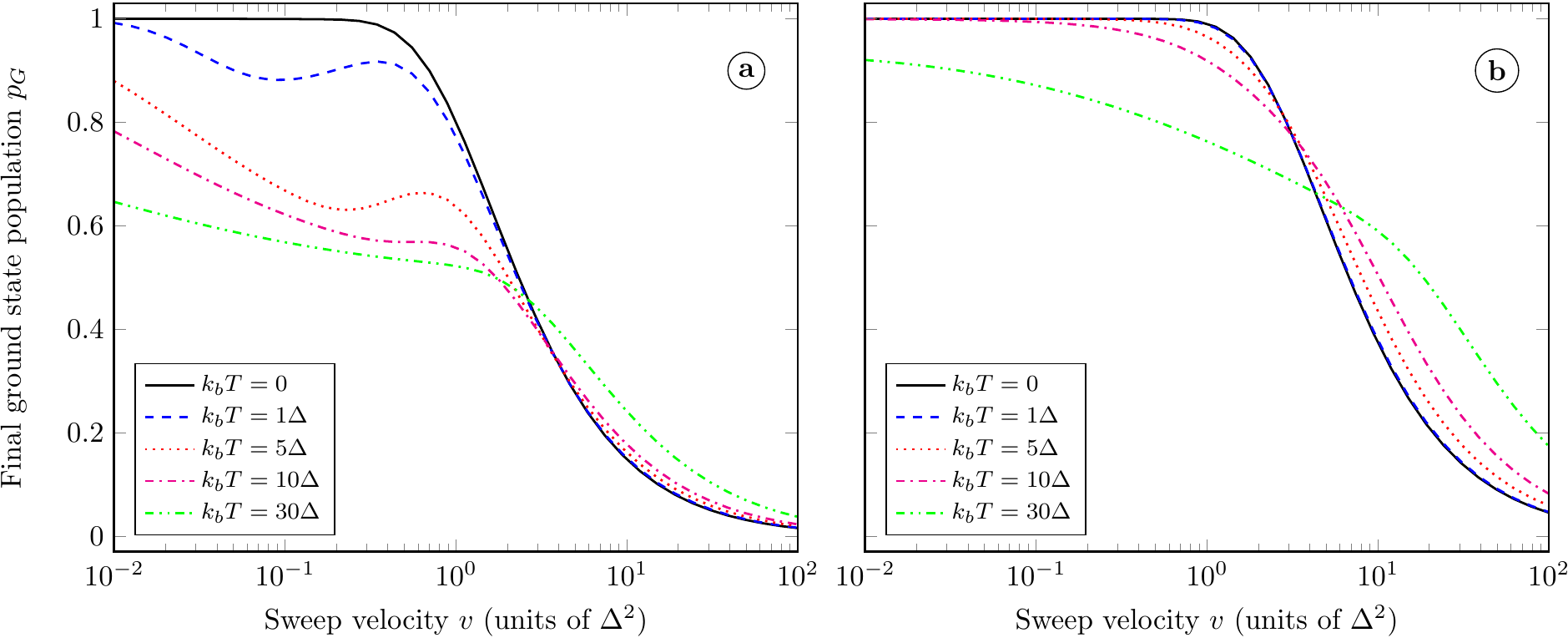}
    \caption{\label{fig:app1}(a) Final ground state population $p_G$ as a function of the sweep velocity $v$ for a $\sz$-only coupling with coupling strength $\alpha_z=5\cdot 10^{-3}$ at different temperatures. (b) Final ground state population for same parameters as in (a) but with additional transverse coupler.}
\end{figure}

\begin{figure}
	\centering
    \includegraphics[width=.8\textwidth]{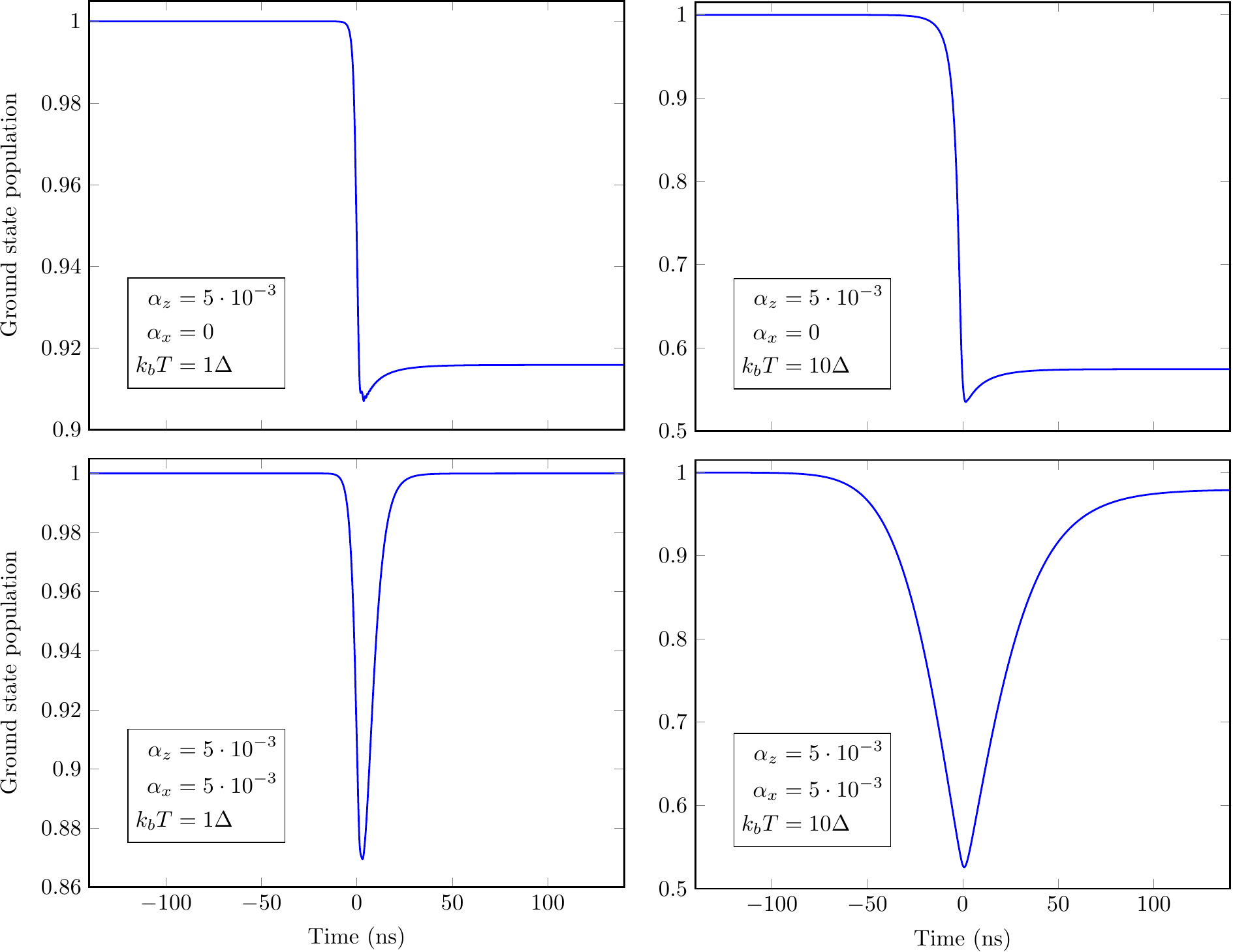}
    \caption{\label{fig:app2}Population of the ground state as a function of time for different parameter settings with sweep velocity $v=0.3\del^2$. As apparent from the plots, an additional transverse coupling does not reduce intermediate excitations. Cooling into the ground state is achieved by relaxation back into the ground state.}
\end{figure}

In addition to the graphics shown in the main article, we want to support the statements by providing further numerical data. For an absolute comparison of how the final ground state population depends on temperature and sweep velocity, please see \reffig{app1}. Our statement that cooling is solely caused by relaxation processes is supported by \reffig{app2}, which depicts the evolution of ground state population for different parameter settings. If the CPW is transversely coupled to the qubit, excitation out of the ground state is not minimized intermediately. Instead, population relaxes back into the ground state after passing the avoided crossing. We find qualitatively identical dynamics for other parameter regimes as well.

\end{document}